**Four wave mixing study on coalescence overgrowth of GaN nanocolumns on sapphire**

Hsiang-Chen Wang [1†], Chun-Ming Yeh [1], Tsung-Yi Tang [2], C. C. Yang [2], T. Malinauskas [3], K. Jarasiunas [3]

[1] Graduate Institute of Opto-Mechatronics, National Chung Cheng University, Chia-Yi, 62102, Taiwan, R.O.C.

[2] Institute of Photonics and Optoelectronics, National Taiwan University, Taipei 10617, Taiwan, R.O.C.

[3] Institute of Applied Research, Vilnius University, LT-10222 Vilnius, Lithuania

[†]Correspondent: hcwang@ccu.edu.tw

**Abstract** Coalescence overgrowth of pattern-grown GaN nanocolumns (NC) on c-plane sapphire substrate with metal organic chemical vapour deposition is demonstrated. The subsequent coalescence overgrowth opens a possibility for dislocation reduction due to the lateral strain relaxation in columnar geometry. We present further growth optimization and innovative characterization of MOCVD layers, overgrown on the columnar structure with varying diameter of colums. Nanoimprint lithography was applied to open circular holes of 250, 300, 450, 600 nm in diameter on the $SiO_2$ layer, deposited on the GaN layer on *c*-plane sapphire template. After growth of ~ 1 µm height GaN nanocolumns, the further coalescence conditions led to an overgrown layer of ~ 2 µm thickness. Transmission Electron Microscopy images showed that in such samples, the small hole for NC growth could stop the propagation of threading dislocation (TD) into an NC. However, new TDs could be formed on the $SiO_2$ masks. On the other hand, in a sample of larger hole size/spacing, TDs could penetrate into NCs. Photoelectrical and optical properties of the overgrown layers and a reference sample were investigated by time-resolved picosecond transient grating (TG) technique and time-integrated photoluminescence (PL). We note 3-4 fold increase of carrier lifetime in the overgrown epilayers when the diameter of colums varied from 250 to 450 nm. This feature is a clear indication of a ~ 4-fold reduced defect density. Less efficient electron trapping in the overgrown GaN is supported by the D=1.6 $cm^2$/s value, typical for bipolar plasma in GaN (while the D value is twice faster in the tamplate, as n<p). Increasing lifetime value with excitation is typical for saturation of electron trapping centers by excess carriers in GaN with dislocation density of ~ $10^9$ $cm^{-2}$ .

# 1. Introduction

The growth of GaN NCs on c-plane sapphire or Si substrate is attractive because the NCs can be dislocation free due to the lateral strain relaxation in the column geometry [1-5]. Also, an InGaN/GaN LED has been grown on NCs to achieve high indium content and high crystal quality [6]. However, for device fabrication, a planar geometry is preferred. Therefore, coalescence overgrowth on such high-crystal-quality GaN NCs becomes an important issue. With coalescence overgrowth, we can prepare low-dislocation-density GaN template for device fabrication. GaN NCs can be grown by MBE and MOCVD with the methods of self-organized growth [7,8], regrowth on a selective mask [9], and catalyst-assisted growth [10, 11]. To implement parallel and vertical NCs with MOCVD, normally patterned growth is preferred. Regularly arranged GaN NCs via patterned MOCVD growth by using interferometric lithography have been demonstrated [12]. NC growth followed by coalescence overgrowth with MBE has also been reported [13]. Recently, MOCVD coalescence overgrowth of MBE-grown self-organized GaN NCs on Si substrate was reported by this group [14]. However, further improvement of the quality of the overgrown layer is needed. The quality of the overgrown layer depends on the quality of column array, including its regularity.

In this paper, coalescence overgrowth of pattern-grown GaN nanocolumns (NC) on c-plane sapphire substrate with metal organic chemical vapour deposition is demonstrated. A quite regular pattern of NC based on nano-imprint lithography is first deposited, followed by a two-dimensional growth procedure, leading to a smooth overgrown layer of significantly higher crystal and optical qualities. Transmission Electron Microscopy images showed that in such a sample, the small hole for NC growth could stop the propagation of threading dislocation (TD) into an NC. In the time-integrated photoluminescence results, a factor of 7-8 stronger PL intensity at room temperature confirmed the increased optical quality of the overgrown layer, what correlates well with the enhanced electrical properties as evidenced from carrier dynamics (decreased nonradiative recombination rate in samples A to C). Origin of strong PL in sample D but short lifetime and the decreased D value can be attributed to role of boundaries between the overgrown domains. In the time-resolved picosecond transient grating results, we note 3-4 fold increase of carrier lifetime in the overgrown epilayers when the diameter of colums varied from 250 to 450 nm. This feature is a clear indication of a ~ 4-fold reduced defect density. Such a GaN template is useful not only for high-quality device fabrication, but also for an easier laser liftoff process in fabricating vertical LEDs.

## 2. Growth Conditions and TEM images of the growth samples

To first prepare the pattern for nanocolumn growth on a GaN template, which consists of a 2μm GaN thin film grown at 1050 °C after the nucleation layer of 40 nm grown at 530 °C on *c*-plane sapphire substrate, an 80 nm $SiO_2$ layer was deposited at 300 °C with plasma-enhanced chemical vapor deposition. Then, nanoimprint lithography was applied to open circular holes of 250, 300, 450, 600 nm in diameter and 500, 600, 900, 1200 nm in distance between the nearest neighboring hole centers with the hexagonal pattern on the SiO2 layer created through reactive ion etching, represented to sample A, B, C, and D, respectively. At the beginning of MOCVD growth, the substrate temperature was set at 1050 °C with a chamber pressure of 100 torr and a V/III ratio of 1100. 5 s after the growth was started, the flow rates of trimethylgallium ( TMGa ) and ammonia were alternatively on/off modulated for nanocolumn growth [12]. During the growth with flow-rate modulation, the flow rates of TMGa and ammonia were 12.5μmol/min and 500 SCCM ( SCCM denotes cubic centimeter per minute at STP ), respectively. The durations of supplying TMGa and ammonia were 20 and 30 s, respectively. GaN nanocolumns of ~ 1μm in height were obtained after a period of 30 min. For coalescence overgrowth, the chamber pressure and V/III ratio were changed into 200 torr and 3900, respectively, with the growth temperature kept at 1050 °C. The continuous flow rates of TMGa and ammonia were 3.5 μmol/min and 1500 SCCM, respectively. Under such growth conditions, the growth rate was ~ 1.3 µm/h. Coalescence overgrowth of 90 min led to an overgrown layer of ~ 2µm in thickness. To demonstrate the improved quality of the coalescence overgrowth on nanocolumns, a sample of the aforementioned GaN template for nanoimprint process was used as the control sample (sample E) for comparison.

Figure 1(a) shows a tilted plan-view SEM image of the hexagonally arranged GaN NCs (sample A), which have the cross-section shape of hexagon. Here, one can see that except a few ones, the sizes of columns are quite uniform. The column heights are about 1 µm and the average cross section size is around 300 nm. The column spacing is fixed at 500 nm by the nano-imprint mold. Along the c-axis, the column cross section is also quite uniform. In Fig. 1(b), we show a cross-section SEM image demonstrating the bottom of a NC. Here, the bottom edges of the $SiO_2$ masks (80 nm in thickness) of slanted walls define the 250-nm hole diameter. The column cross-section size (300 nm) is larger than the hole diameter. Figures 2 (a) and (b) show the typical TEM images of samples A and D, respectively. It is noted that the top portion of sample A was removed during TEM sample preparation such that the illustrated overgrowth layer in Fig. 2(a) is only a few hundred nm in thickness. Such a result does not affect our important observation, which is focused at the NC layer. In the NC layer, the NC gaps have been filled up during the overgrowth stage. In Fig. 2(a), a few circles are shown to draw the attention for the important points. Circle 1 shows the termination of a TD at the emergence of an NC. Circle 2 shows the stop of a TD by the $SiO_2$ mask. In this figure, basically there is no dislocation propagating into an NC. However, voids and new TDs are formed above the masks. As indicated by circles 3-7, at least a TD is formed on the top of each mask region. Some of such TDs propagate into the overgrown layer, such as that indicated by circle 5. Other TDs are terminated through bending or SF formation, such as those indicated by circles 3, 4, 6, and 7. It is believed that the overgrowth mechanism in this case of small NC and small mask area is different from that under the micron-scale condition of the ELOG technique.

On the other hand, in the case of 600-nm hole size (sample D), as shown in Fig. 2(b), plenty TDs can propagate into NCs, as indicated by circles 1-3, and continue the propagation into the overgrown layer, as indicated by circles 8 and 9. It is noted that there is an abrupt change of sample thickness (formed during TEM sample preparation) around circle 1 such that the TDs suddenly disappear after they penetrate into the layer above the mask. In this figure, the TDs stopped by the masks are indicated by circles 4-7. In this sample, the mask area is large such that the overgrowth process is similar to that in the conventional ELOG technique. With such a lateral overgrowth process, the TD density in a mask region is significantly lower than that in an NC. Therefore, we can speculate that in an overgrowth sample of a smaller hole size/spacing, the TDs in the overgrown layer originate mainly from the mask regions. In this situation, the small hole size terminates TD propagation; however, the small spacing (mask size) leads to the formation of new TDs. On the other hand, in a sample of a larger hole/spacing, the TDs in the overgrown layer originate mainly from the NC regions. In this situation, the large hole size allows the penetration of TDs from the template; however, the large spacing allows more effective lateral overgrowth and leads to relatively fewer new TDs. The lateral overgrowth results in the formation of SFs, as exemplified by those indicated by the two dashed arrows. It is noted that the TDs indicated by circle 10 belong to an NC behind the mask of circle 4. They appear in Fig. 2(b) because of the aforementioned abrupt change of TEM sample thickness.

## 3. Photoluminescence and Four wave mixing results

Figure 3 shows the temperature-dependent integrated PL intensities of the five samples. Because the absorption coefficient of GaN at 325 nm is around $1 \times 10^5$ cm$^{-1}$ [15], the PL measurement reflects the optical quality of a layer of ~100 nm in thickness near the surface of the overgrown layer. One can see that the spectral distributions of the overgrown layer and the control sample are similar. However, that of the NCs sample is significantly red shifted. The two peaks in the PL spectra of all the five samples (around 362 and 368 nm in the overgrowth and control samples) are formed because of the existence of a reabsorption feature between the two spectral peaks [16].

Room temperature carrier dynamics in GaN was investigated using light-induced transient grating (LITG) technique [17]. The pulses from a Nd:yttritium–aluminum–garnet laser at $\lambda = 355$ nm (h$\nu$ = 3.49 eV) intersected at the sample and created an interference pattern with fringe spacing $\Lambda \approx \lambda/\sin(\theta)$, where $\theta$ is the angle between the incident beams. The free-carrier grating $N(x) = \Delta N [1+ \cos(2\pi x/\Lambda)]$ was created at the very surface by interband absorption since h$\nu$ > $E_g$ = 3.39 eV, and the sample excitation depth was close to an inverse absorption coefficient $1/\alpha \approx 150$ nm. Carrier diffusion to the sample depth significantly expanded the excited area [18]. A delayed probe beam at longer wavelength, $\lambda = 1064$ nm, to which the sample is transparent, monitored the modulation of excess electron-hole density $\Delta N$ as a function of time via the time-varying diffraction efficiency of the grating: $\eta(t) \propto \Delta N(t)^2$. The period of transient grating was changed in a range from 3 to 7 $\mu$m by varying the angle $\theta$, thus providing an unique advantage to measure in-plane carrier diffusion and lifetime. The modulated part of free carrier concentration $\Delta N$ decays with characteristic grating decay time $\tau_G$ governed by carrier recombination and in-plane diffusion with coefficient D [17,18]:

$$\frac{1}{\tau_G} = \frac{1}{\tau_R} + \frac{1}{\tau_D} = \frac{1}{\tau_R} + \left(\frac{2\pi}{\Lambda}\right)^2 D \qquad (1)$$

Therefore, measurement of grating decay times $\tau_G$ at different grating periods $\Lambda$ allows independent determination of carrier lifetime $\tau_R$ and diffusion coefficient D. Previous studies of carrier dynamics in GaN layers were focused mainly on TD-density dependent nonradiative recombination time. These studies were summarized in Ref [17] and revealed variation of $\tau_R$ from ~50-100 ps in MOCVD grown layers on Si (with $N_{TD} = 10^9 - 10^{10}$ cm$^{-2}$) up to ~1 ns in standard grown layers on sapphire (with ~mid-$10^8$ cm$^{-2}$ density). Specific treatment of the Si/N buffer layer led to NC growth of GaN layers [19] with TD density of

~$7.10^7$ cm$^{-2}$ level and nonradiative carrier lifetime of few ns [17]. Advanced HVPE growth allowed increasing carrier lifetime up to 15 ns [17]. Moreover, determination of $\tau_R$ and D values by LITG technique provided the carrier diffusion length, which directly revealed the structuraquality of materials. Consequently, control of photoelectrical parameters in NC layers gown at various conditions may he helpful for optimization of the growth process.

The measurements were carried out at RT at various excitations energy densities I. Decay of LITG recorded in the layer overgrown on the 250 nm columns is shown in Fig. 4. The decay kinetics were nearly exponential and faster for smaller grating periods, in accordance with Eq (1). Plotting the dependence of grating decay rate $1/\tau_G$ vs inverse grating period $\Lambda^{-2}$ [20], the values of $\tau_R$ and D were determined. Similarly, we measured the grating decay rates in NC layers, overgrown on different diameter nanocolumns as well in the reference sample E. The determined parameters are given in Tables 1. The data indicated that the lifetime increased in the overgrown layers with respect to reference template. This tendency is presented in Fig 5. We note that the samples A and B (overgrown on the small diameter columns) had the longest carrier lifetimes and typical D values for GaN [18]. Complementary measurements of time integrated PL intensity at 325 nm excitation (Fig 3) pointed out to the increased radiative recombination component due to de creased nonradiative recombination rate in overgrown layers.

It is worth to note the presence of high density of defects in the used template that might also influence the quality of the overgrown layers. The observed very short carrier lifetime of ~150 ps in the template and its ~3-fold increase with excitation (see Fig.6) reflect the high TD density and partial saturation of electron trapping centers by injected carriers. The similar feature (three-fold increase of carrier lifetime from 240 to 720 ps) was earlier observed in the interface region of a standard 3-4 μm thick MOCVD layer with $N_{TD}$ in mid-$10^8$ cm$^{-2}$ [21]. In the given template (sample E) and in the overgrown layers A, B this effect was present on the front side of the GaN epilayer.

## 4. Conclusion

In conclusion, by complementary optical techniques we confirmed the improved electrical, optical, and thus structural parameters of NC layers on different diameter columns. Carrier lifetime of 500 ps and diffusion length $L_D$ of 0.28 μm was found optimal for the layers overgrown on 250 nm columns. We note 3-4 fold increase of carrier lifetime in the overgrown epilayers when the diameter of colums varied from 250 to 450 nm. This feature is a clear indication of a ~ 4-fold reduced defect density. Less efficient electron trapping in the overgrown GaN is supported by the D=1.6 cm$^2$/s value, typical for bipolar plasma in GaN (while the D value is twice faster in the tamplate, as n<p). Increasing lifetime value with excitation is typical for saturation of electron trapping centers by excess carriers in GaN with dislocation density of ~ $10^9$ cm$^{-2}$ .

## 5. Acknowledgement

This research was supported by National Science Council, The Republic of China, under the grants of NSC 97-2120-M-002-005, NSC 97-2622-E-002-011-CC1, NSC 96-2628-E-002-044-MY3, NSC 97-2221-E-002-044, NSC 97-2221-E-194-008 by US Air Force Scientific Research Office under the contracts AOARD-07-4010 and AOARD-09-4117, by the Epistar Corporation, Taiwan and European Commission Contract No. MRTN-CT-2006-035735.

**List of table and figure captions**

Table 1 Carrier lifetime $\tau_R$ (ps) and diffusion coefficient D (cm$^2$/s) in MOCVD layers, overgrown on columnar structure with smaller diameter of columns (A, B, C, D) with respect to the reference template (E).

Figure 1. (a) A tilted plan-view SEM image of hexagonally arranged NCs, which have the cross-section shape of hexagon. Except a few relatively larger columns, the sizes and heights of those columns are quite uniform. (b) A cross-section SEM image showing the bottom of a NC. The bottom edges of the SiO$_2$ masks (100 nm in thickness) of slanted walls define the 250-nm hole diameter. The column cross-section size is around 300 nm.

Figure 2. (a) Cross-sectional TEM image of sample A. A top portion of the sample was removed during TEM sample preparation. Various TD features are indicated with circles. (b) Cross-sectional TEM image of sample D. Various TD features are indicated with circles.

Figure 3. Photoluminescence spectra in the overgrown samples (A, B, C, D) and reference sample (E)

Figure 4. Diffraction kinetics in NC sample A for three grating periods provided $\tau_R$ = 500 ps and D= 1.65 cm$^2$/s values.

Figure 5 Diffraction kinetics in the set of NC samples (A to D) and in the reference template.

Figure 6 Diffraction kinetics in the reference template (sample E) at different excitation energy densities I.

**Tables**

Table 1 Carrier lifetime $\tau_R$ (ps) and diffusion coefficient D (cm$^2$/s) in MOCVD layers, overgrown on columnar structure with smaller diameter of columns (A, B, C, D) with respect to the reference template (E).

| $I_0$ [mJ/cm$^2$] | A (250 nm) | | B (300 nm) | | C (450 nm) | | D (600 nm) | | E (reference) | |
|---|---|---|---|---|---|---|---|---|---|---|
| | $D_a$ | $\tau_R$ | $D_a$ | $\tau_R$ | $D_a$ | $\tau_R$ | $D_a$ | $\tau_R$ | $D_a$ | $\tau_R$ |
| 1.5 | 1.65 | 501 | 1.55 | 359 | 0.44 | 672 | <1 | 134 | 3.04 | 150 |
| 3.0 | 1.62 | 509 | 1.66 | 538 | 0.66 | 768 | <1 | 264 | 3.07 | 220 |
| 5.4 | 1.25 | 580 | 1.72 | 763 | 0.66 | 774 | 1.26 | 356 | 4.3 | 390 |
| 8.3 | 1.94 | 890 | 1.8 | 791 | 1 | 770 | 1.77 | 400 | 3.26 | 480 |

**Figures**

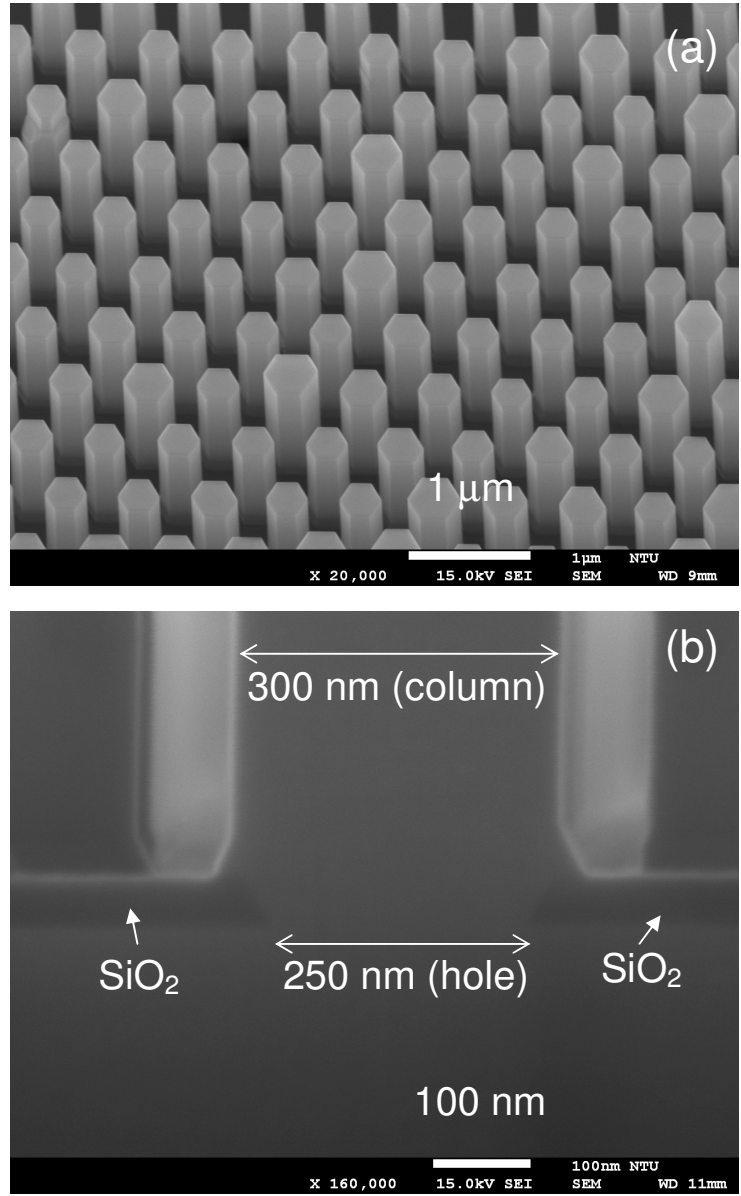

Figure 1. (a) A tilted plan-view SEM image of hexagonally arranged NCs, which have the cross-section shape of hexagon. Except a few relatively larger columns, the sizes and heights of those columns are quite uniform. (b) A cross-section SEM image showing the bottom of a NC. The bottom edges of the $SiO_2$ masks (100 nm in thickness) of slanted walls define the 250-nm hole diameter. The column cross-section size is around 300 nm.

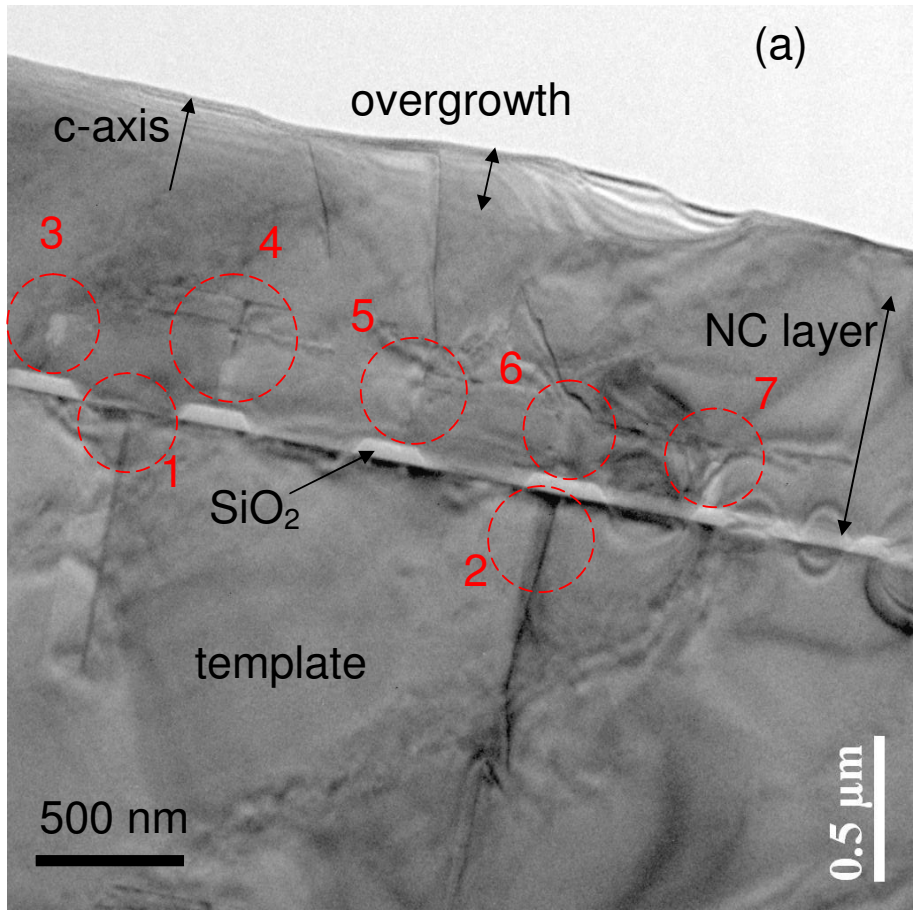

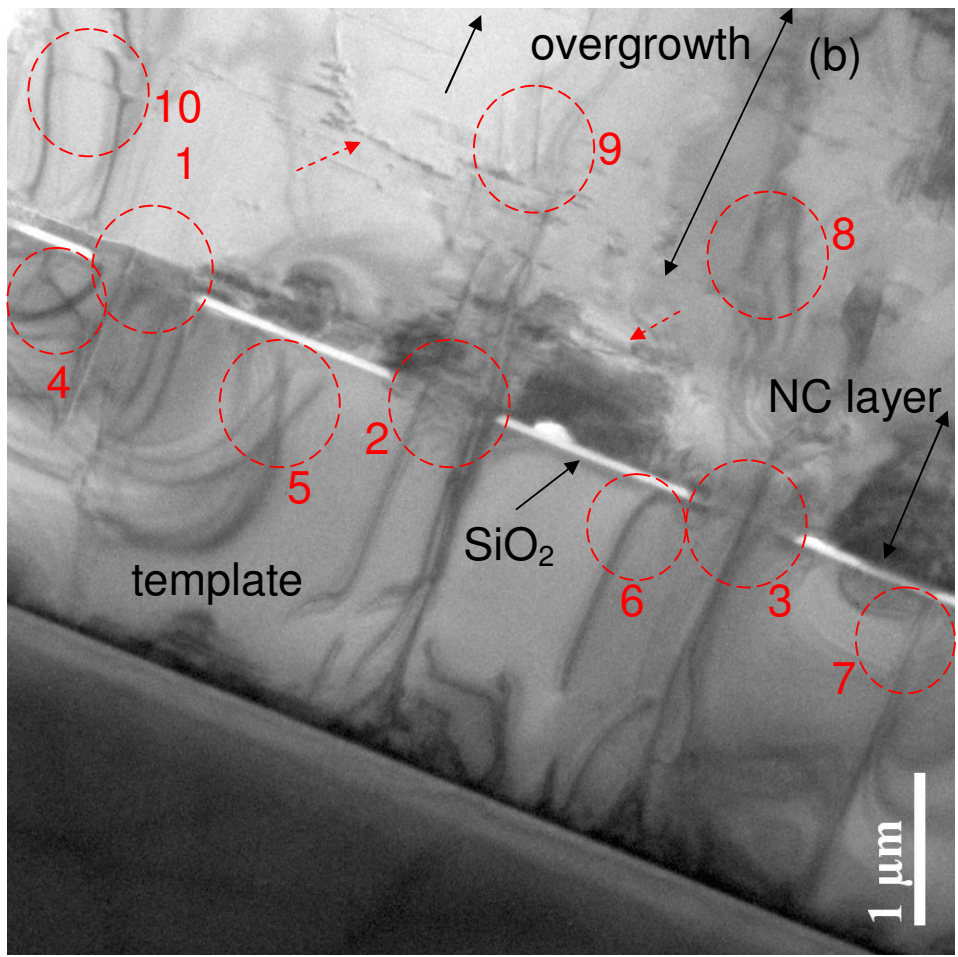

Figure 2. (a) Cross-sectional TEM image of sample A. A top portion of the sample was removed during TEM sample preparation. Various TD features are indicated with circles. (b) Cross-sectional TEM image of sample D. Various TD features are indicated with circles.

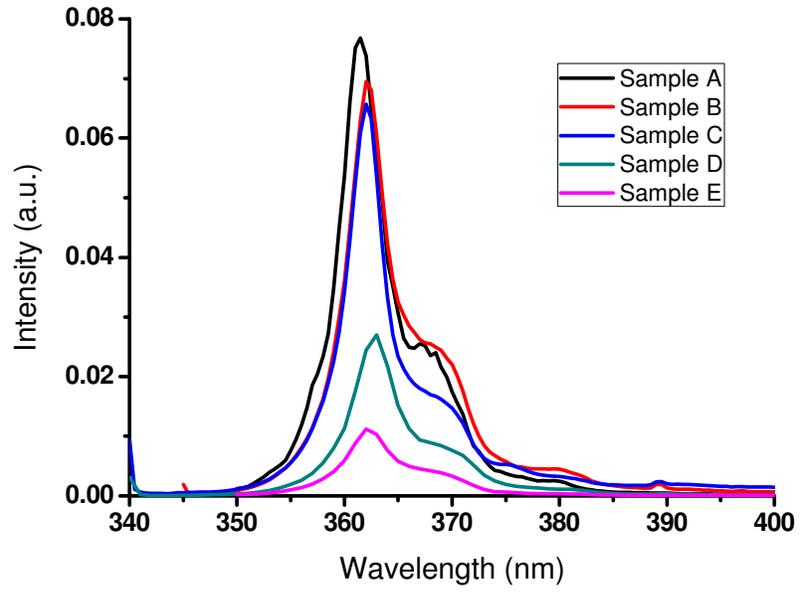

Fig.3 Photoluminescence spectra in the overgrown samples (A, B, C, D) and reference sample (E).

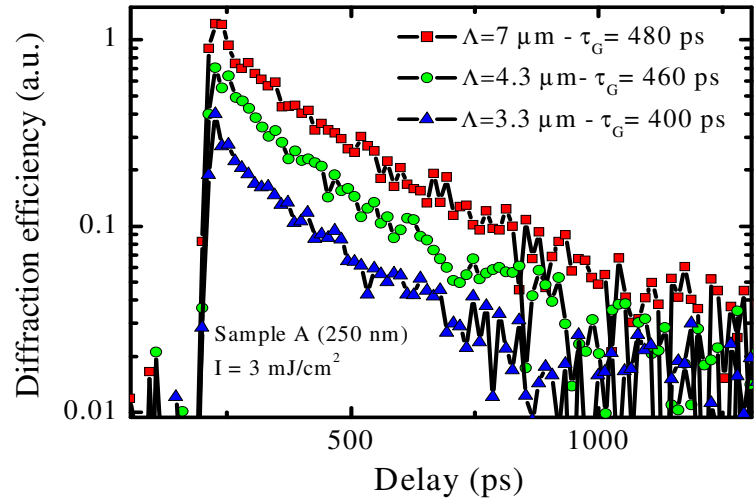

Figure 4. Diffraction kinetics in NC sample A for three grating periods provided $\tau_R$ = 500 ps and D= 1.65 cm$^2$/s values.

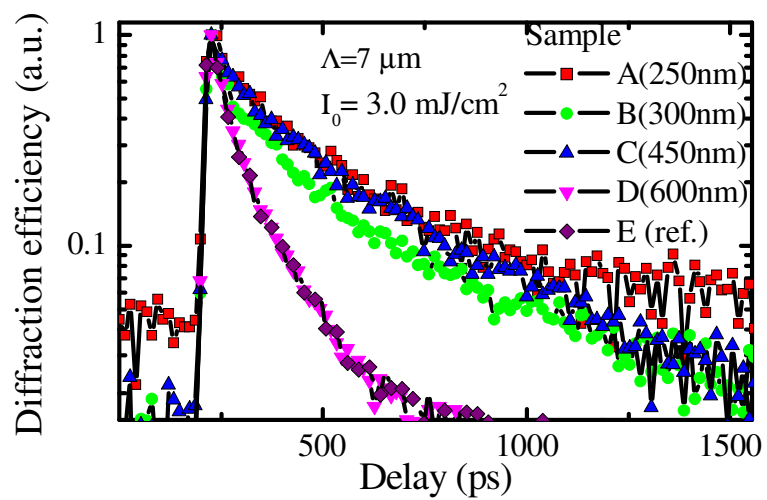

Fig. 5 Diffraction kinetics in the set of NC samples (A to D) and in the reference template.

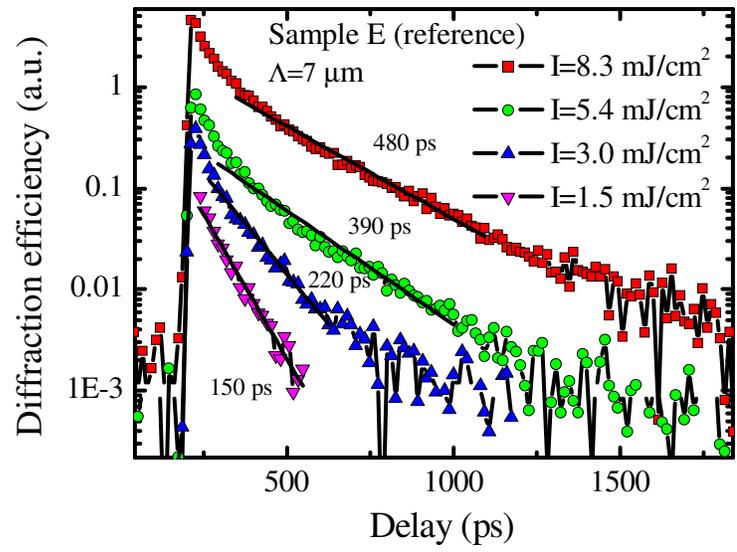

Fig. 6 Diffraction kinetics in the reference template (sample E) at different excitation energy densities I.